\pdfoutput=1
\documentclass[12pt,a4paper]{article}
\usepackage{jheppub}
\usepackage[T1]{fontenc} 
\usepackage{graphicx}
\usepackage{epsfig}
\usepackage{rotating}
\usepackage{amssymb}
\usepackage{dsfont}
\usepackage{psfrag}
\usepackage{amsmath,euscript,array,mathrsfs}
\usepackage{bbold}
\usepackage{epsf}
\usepackage{slashed}
\usepackage{cancel}
\usepackage{float}
\usepackage{subcaption}
\usepackage{enumerate}

\usepackage{caption,subcaption,wrapfig}
\usepackage[colorlinks=true,linktocpage=true,linkcolor=blue,citecolor=blue]{hyperref}

\usepackage{tikz}
\usetikzlibrary{decorations.markings,decorations.pathmorphing}

\newcommand{\be}{\begin{equation}}
\newcommand{\ee}{\end{equation}}
\newcommand{\beqs}{\begin{eqnarray}}
\newcommand{\eeqs}{\end{eqnarray}}

\newcommand{\dd}{\mathrm{d}}

\makeatletter
\gdef\@fpheader{}
\makeatother

\begin{document}

\preprint{HIP-2025-22/TH}

\title{Testing the effective action approach to bubble nucleation in holography}

\author[a,c]{Oscar Henriksson,}
\author[b,c]{Niko Jokela,}  
\author[b,c]{Xin Li}

\affiliation[a]{Faculty of Science and Engineering, {\AA}bo Akademi University,\\ Henrikinkatu 2, FI-20500 Turku, Finland}
\affiliation[b]{Department of Physics, University of Helsinki,\\ P.O.~Box 64, FI-00014 University of Helsinki, Finland}
\affiliation[c]{Helsinki Institute of Physics,\\ P.O.~Box 64, FI-00014 University of Helsinki, Finland}

\emailAdd{oscar.henriksson@abo.fi}
\emailAdd{niko.jokela@helsinki.fi}
\emailAdd{xin.z.li@helsinki.fi}

\abstract{The nucleation of bubbles during a first-order phase transition has recently been explored using holographic duality, which can provide an important complement to standard perturbative methods. These computations typically require finding static and spatially inhomogeneous saddle points, known as critical bubbles, which correspond in the gravitational dual to solutions of nonlinear partial differential equations. A computationally simpler alternative is to use the gravitational dual to derive the effective action of the boundary theory in a derivative expansion, and then solve the resulting lower-dimensional equations of motion. Once the effective action, typically truncated at two derivatives, is obtained, the holographic theory can be set aside, and bubble solutions can be found from ordinary differential equations. In this paper, we test this approach in a simple holographic setup: a scalar field in the probe limit in a black brane background, with nonlinear multi-trace boundary conditions. We compute critical bubble solutions both from the effective action and by solving the scalar field equation of motion directly in the gravity theory, and find good agreement between the two methods.}

\maketitle

\section{Introduction}

First-order phase transitions (FOPTs) appear across many areas of physics, from simple thermodynamic systems to models of the early universe. A key feature of such transitions is the nucleation of bubbles, localized field configurations that mediate the decay of a metastable phase into a stable one. In path-integral computations, these bubbles appear as nonperturbative saddle points, and are typically studied by solving the equation of motion (EoM) derived from an effective action for the order parameter~\cite{LANGER1967108,LANGER1969258,Coleman:1977py,Linde:1980tt,Linde:1981zj}. Physical quantities can then be extracted; for example, the nucleation rate per volume is proportional to $e^{-B}$, where $B$ is computed from the action of the bubble solution. Computational studies of bubble nucleation are often challenging, especially at strong coupling.

In gauge/gravity or holographic duality, strongly coupled quantum field theories (QFTs) in $d$ spacetime dimensions can be studied via dual gravitational models in $d+1$ dimensions. This framework enables access to otherwise challenging regimes such as bubble nucleation, see \cite{Bigazzi:2020phm,Ares:2020lbt,Li:2020ayr,Bigazzi:2021ucw,Bea:2021zsu,Henriksson:2021zei,Ares:2021ntv,Zhu:2021vkj,Ares:2021nap,Wang:2023lam,Henriksson:2024hsm,Janik:2025uqa} for a non-exhaustive sample. However, the construction of inhomogeneous configurations in gravitational theories, including critical bubbles, typically requires solving coupled nonlinear partial differential equations (PDEs), often with nontrivial boundary conditions. This makes direct numerical treatment of such processes challenging.

As an alternative to directly constructing inhomogeneous solutions in the gravity theory, one can instead use the gravitational description to derive an effective action for the order parameter\footnote{Note that we use the term \textit{order parameter} despite the fact that no spontaneous symmetry breaking takes place in the phase transitions studied below.} $\psi$ in the boundary theory~\cite{Hertog:2004ns,Hertog:2005hu,Papadimitriou:2007sj,Kiritsis:2012ma,Kiritsis:2014kua,Ares:2021ntv,Ares:2021nap}. This effective action can be organized in a derivative expansion,
\begin{equation}\label{eq:effaction}
    \Gamma[\psi] = \int \dd^d x \left[ -V(\psi) - \frac{1}{2}Z(\psi)\partial_\mu\psi \partial^\mu\psi + \ldots \right] ,
\end{equation}
where the ellipses denote terms with four or more derivatives with respect to the field theory directions and $\mu,\nu=0,1,\ldots,d$. As we review in detail later, the holographic dictionary can be used to extract the functions $V(\psi)$ (the effective potential) and $Z(\psi)$ by analyzing the properties (including linearized fluctuations) of \emph{static and homogeneous} gravity solutions.\footnote{Higher-derivative terms can also be computed along the same lines, although the complexity and the number of terms quickly increases.} Then, the EoM resulting from this effective action can be solved to find critical bubbles and other non-trivial solutions of interest. This approach only requires us to solve a number of ordinary differential equations (ODEs), a much less demanding task which can typically be accomplished by standard shooting methods. In \cite{Ares:2021ntv,Ares:2021nap}, bubble nucleation was studied in this way in a simple holographic gravity-scalar theory.

While the effective action approach greatly simplifies the study of bubble nucleation in holography, one could worry that the truncated derivative expansion introduces errors in the bubble solutions which could potentially affect the physics. The goal of this paper is to test the accuracy of this approach by explicitly constructing bubble solutions directly in the bulk gravitational theory and comparing them to those obtained from the two-derivative effective action. The gravitational construction involves solving a nonlinear PDE for the scalar field profile in the bulk, which should account for all orders in derivatives in the boundary field theory. Our results show excellent agreement between the two methods across a range of parameter values, both in the bubble profiles and in derived quantities such as the on-shell action. This supports the idea that the effective action approach, even when truncated at second order in derivatives, captures the essential features of bubble nucleation in this class of holographic models.

We note that this paper is only concerned with efficiently computing critical bubble solutions in a holographic theory. Important follow-up questions remain, such as computing the fluctuation determinants that determine the prefactor multiplying $e^{-B}$ in the nucleation rate using the gravitational theory; we leave these for future work.

The rest of the paper is organized as follows. In Sec.~\ref{sec:model}, we present the holographic model, including the probe limit and boundary deformation. In Sec.~\ref{sec:bubble_eff}, we construct the dual field theory effective action up to two derivatives, and solve for critical bubbles using the resulting EoM. Sec.~\ref{sec:bubble_gr} presents the direct construction of bubbles from the gravitational equations. We compare the results and discuss the level of agreement in Sec.~\ref{sec:comparison}. We conclude in Sec.~\ref{sec:discussion} with a discussion of broader applications and future directions.

\section{Gravitational model}\label{sec:model}

In order to test the effective action approach, we choose to study one of the simplest possible holographic bottom-up models which displays FOPTs and the associated bubble solutions. The model is a four-dimensional\footnote{The choice of dimension is mainly for technical convenience; combined with the choice of mass detailed below, it gives a simple boundary falloff for the scalar field while allowing us to impose suitable boundary conditions. We do not expect our results to be sensitive to this choice.} bottom-up gravity-scalar theory with a simple quartic scalar potential and a negative cosmological constant:
\begin{equation}\label{eq:bulkAction}
S=\frac{1}{2\kappa_{4}^{2}}\int\mathrm{d}^{4}x\sqrt{-g}\left[\mathcal{R}+\frac{6}{L^2}-\frac{1}{N} \left(\partial_M\phi\partial^M\phi+m^{2}\phi^{2}+\frac{1}{4}\phi^{4}\right)\right] \ .
\end{equation}
Here, $\cal R$ is the scalar curvature, $g$ is the determinant of the spacetime metric, $\phi$ is a scalar field with mass $m$, $L$ is the anti-de Sitter space (AdS) radius, $\kappa_4^2=8\pi G_4$ is proportional to the Newton constant $G_4$, and indices $M=0,1,2,3$. We have included a dimensionless factor $1/N$ in front of the scalar part of the Lagrangian, since we want to work in the probe limit where $N$ is large and the scalar field does not backreact on the metric. In the purported field theory dual, we can think of this scalar as describing a smaller ``flavor'' sector, which couples to a larger ``color'' sector --- described by the gravitational part --- which dominates the thermodynamics. We then posit that $N$ effectively acts as the ratio of degrees of freedom in the color and flavor sectors. As we will see, the phase transition will be entirely caused by the flavor sector, \textit{i.e.}, the scalar field.

This theory admits AdS-Schwarzschild solutions where $\phi=0$. Using scaling symmetries of the action to set $L=1$, and working in ingoing Eddington--Finkelstein coordinates, these take the form
\begin{align}
\dd s^2 & =\frac{1}{u^2}\left(-f(u) \mathrm{d} t^2-2\mathrm{d} t \mathrm{d} u+\mathrm{d} \vec{x}^2\right) \\
f(u) & =1-\left(\frac{u}{u_H}\right)^3,
\end{align}
where horizon and  conformal boundary locate at $u=u_H$ and $u=0$, respectively. The Hawking temperature for these solutions is 
\begin{equation}
	T=\frac{3}{4\pi u_H} \ .
\end{equation}
Throughout this paper we measure everything in units of $T$, which can be accomplished by setting $u_H=3/4\pi$. 

Without any further ingredients, the above theory is too simple to display a phase transition, particularly in the probe limit. One attempt to improve on this would be to introduce a more complicated potential for the scalar field \cite{Bea:2018whf}. We will instead follow \cite{Hertog:2004ns,Ares:2021nap,Ares:2021ntv}, keeping the gravity Lagrangian simple and obtaining a FOPT by imposing nonlinear Robin boundary conditions on the scalar field. In the dual field theory, these boundary conditions implement so-called multi-trace deformations \cite{Witten:2001ua}. The name multi-trace comes from the fact that typical top-down holographic field theories are large-$N$ gauge theories, whose simplest gauge-invariant local operators are (single) traces of matrix-valued fields. Products of such quantities are called multi-trace operators. In the large-$N$ limit, deforming the action with such operators changes the effective potential in a simple way \cite{Papadimitriou:2007sj}: If the $n$th power of a single-trace operator $\mathcal{O}$ is added to the action, $S\to S+g_n \mathcal{O}^n$, the effective potential is deformed as $V(\psi)\to V(\psi)+g_n \psi^n$, where $\psi$ can be thought of as the expectation value of the operator. It is thus simple to relate the multi-trace boundary conditions directly to the resulting phase diagram; compare this to the bulk scalar potential, whose impact on the phase diagram is much less clear.

To add multi-trace deformations, the single-trace operator dual to the scalar $\phi$ must have scaling dimension $\Delta\le 3/2$, otherwise the deformations will be irrelevant. From the standard holographic dictionary, this forces the scalar field to have a mass in the range allowing for alternate quantization \cite{Breitenlohner:1982bm}, which in four spacetime dimension restricts us to $-9/4 \le m^2 \le -5/4$. A convenient choice is $m^2=-2$; with this choice, the dual single-trace operator has dimension 1, and we can include a relevant double-trace deformation and a marginal triple-trace deformation. (We choose not to turn on the more common single-trace deformation simply because this is not needed for our purposes.) As mentioned above, this modifies the effective potential by a quadratic and a cubic term:
\begin{equation}\label{eq:effPotDeformation}
    V(\psi) \to V(\psi) + \frac{g_2}{2} \psi^2 + \frac{g_3}{3} \psi^3 \ .
\end{equation}
By tuning the coefficients $g_2$ and $g_3$ of these deformations, the theory can be made to realize FOPTs, as we show in the next section.

Let us now briefly explain the necessity of the quartic term in the bulk potential for the scalar field $\phi$ in (\ref{eq:bulkAction}). Without this term, the EoM for the scalar field in the probe limit will be linear, making the undeformed effective potential quadratic in $\psi$. To engineer a FOPT, the simplest modification is to turn on an asymmetric $\psi^3$-deformation. This, however, would lead to a potential unbounded from below. Having nonlinearity in the scalar EoM (\ref{eq:bulkAction}) leads to a more complicated effective potential, but which remains bounded from below at least for small values of $g_3$.

With the choice $m^2=-2$, the scalar EoM is
\begin{equation}\label{eq:scalarEoM}
	\nabla_M \nabla^M \phi+2\phi-\frac12 \phi^3=0 \ ,
\end{equation}
where $\nabla_M$ is the four-dimensional covariant derivative. By inspecting the EoM one finds that the scalar falloff near the conformal boundary goes as
\begin{equation}
	\phi=\phi^- u+\phi^+ u^2+\ldots \ .
\end{equation}
In App. \ref{app:holorenorm}, we carefully work through the holographic renormalization of this theory, including counterterms which implement alternate quantization and multi-trace deformations. We then find that the expectation value of the dual operator is given by $\psi=-\phi^-$, and that the exact form of the boundary condition we want to impose is
\begin{equation}\label{eq:nonlinearBC}
    \phi^+ = J - g_2\phi^- + g_3(\phi^-)^2 \ ,
\end{equation}
where $J$ is the source (or coupling) for the single-trace operator. We find that working in terms of the rescaled field $\Phi(x)\equiv \phi(x)/u$ leads to some simplifications in the numerics, so we do this from now on. The falloff of the field then becomes
\begin{equation}\label{eq:Phi_bdry_exp}
	\Phi=\phi^- +\phi^+ u+\ldots \ .
\end{equation}

\section{Bubbles from the field theory effective action}\label{sec:bubble_eff}

In this section, we use holographic duality to build the $(2+1)$-dimensional field theory effective action up to second order in derivatives, and then use this to map out the phase diagram and find critical bubble solutions. We begin by a quick review of some properties of the theory effective action, see, \textit{e.g.}, \cite{Schwartz:2014sze} for more details.

In general, for a field theory with field(s) $\Psi$, the quantum (or 1PI) effective action can be defined by starting from the generating functional $W[J]=-i\log{Z[J]}$, where $Z[J]$ is the path integral with an external source $J(x)$, and Legendre transforming:
\begin{equation}
    \Gamma[\psi] = W[J] - \int \dd^3x\, \psi(x)J(x) \ .
\end{equation}
Here, $\psi(x)=\frac{\delta W[J]}{\delta J(x)}$ (sometimes referred to as the \emph{mean} or \emph{classical field}) corresponds to the expectation value of $\Psi$ in the presence of the source $J(x)$. It follows that
\begin{equation}\label{eq:effActionVariation}
    \frac{\delta \Gamma[\psi]}{\delta\psi(x)} = -J(x) \ .
\end{equation}
The full effective action is generally a very non-local object, but can admit a derivative expansion around a homogeneous state. Since we will be working at non-zero temperature where Lorentz invariance is broken, we separate time and space derivatives, writing
\begin{equation}\label{eq:effectiveAction}
    \Gamma[\psi] = \int \dd^3 x \left[ -V(\psi) + \frac{1}{2}Y(\psi)(\partial_t\psi)^2 - \frac{1}{2}Z(\psi)\nabla\psi\cdot\nabla\psi + \ldots \right] \ ,
\end{equation}
where $V(\psi)$ is the effective potential, $Y(\psi)$ and $Z(\psi)$ are non-canonical kinetic terms, and the ellipses represent terms with four or more derivatives. Since the critical bubble solutions we are after are static solutions, we will set the time derivatives to zero from now on. Taking the functional derivative as in Eq.~(\ref{eq:effActionVariation}) then gives
\begin{equation}\label{eq:quantumEoM}
    -J(x) = \frac{\delta\Gamma[\psi]}{\delta\psi} = -V'(\psi) + \frac{1}{2}Z'(\psi)\nabla\psi\cdot\nabla\psi + Z(\psi)\nabla^2\psi + \ldots \ .
\end{equation}
With the source $J(x)$ set to zero, this gives the (quantum) EoM for $\psi$. This is what we will solve later to obtain critical bubble solutions. First, however, we will explain how holography determines the functions $V(\psi)$ and $Z(\psi)$.

\subsection{Determining the effective potential}

To compute the effective potential $V(\psi)$, we note that for static and homogeneous solutions, (\ref{eq:quantumEoM}) reduces to $J(\psi)=V'(\psi)$, which can be integrated to give
\begin{equation}\label{eq:effPotentialIntegral}
    V(\psi) = V(0) + \int_0^\psi \dd\psi'\, J(\psi') \ .
\end{equation}
Since the holographic dictionary relates the near-boundary falloff of the field $\phi$ to $\psi$ and $J$, we can construct the function $J(\psi)$ from the gravity theory, which can then be integrated to find the effective potential (up to the constant $V(0)$ which we can set to zero).

\begin{figure}[t!]
\centering
\subfloat[The source $J(\psi)$.\label{fig:phi_psi}]{\includegraphics[width=0.45\textwidth]{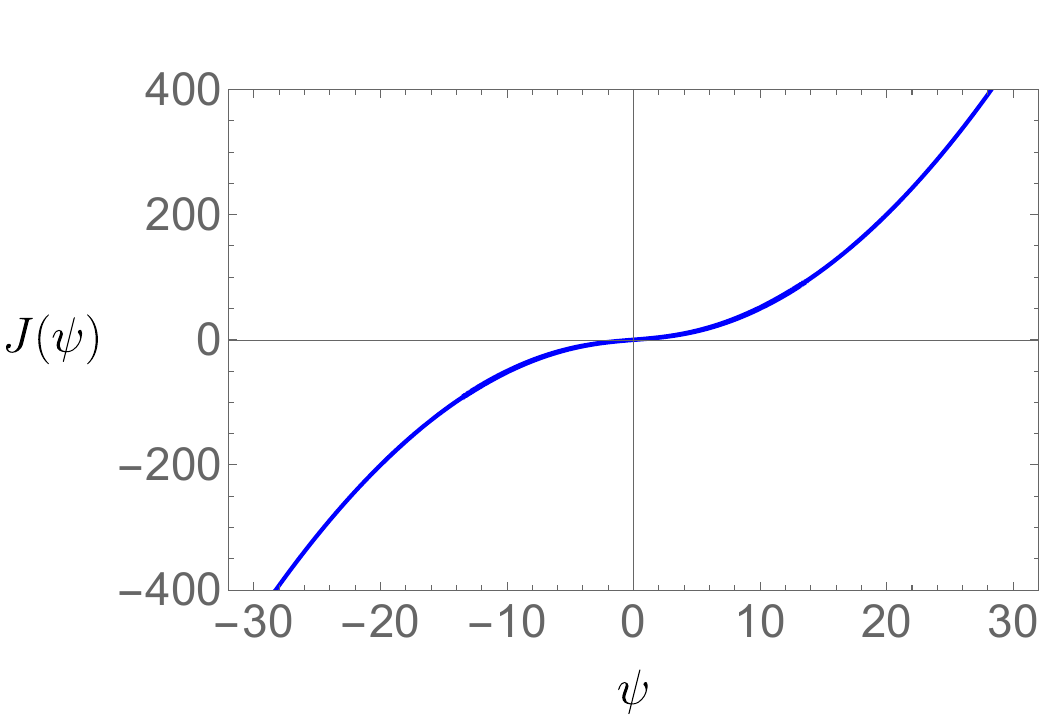}}
	\qquad
\subfloat[The effective potential $V(\psi)$.\label{fig:V_undeformed}]{\includegraphics[width=0.48\textwidth]{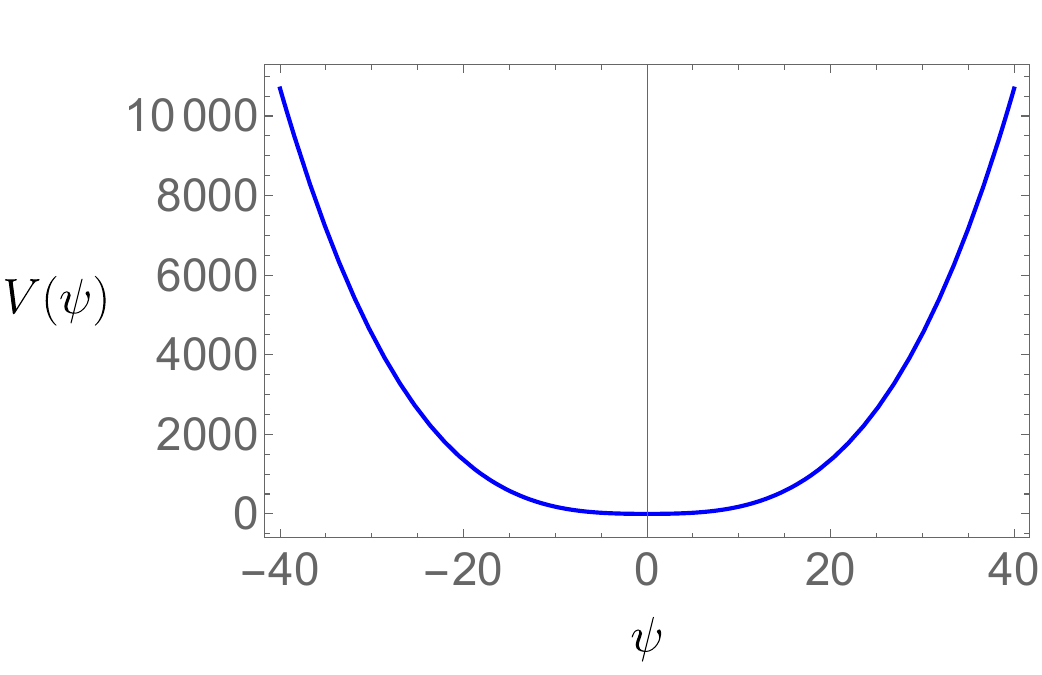}}
\caption{Results for the undeformed system.}
\end{figure}

To do this, we only need to consider solutions to the EoM for our redefined field $\Phi$ which are static and homogeneous in the field theory directions, meaning that we take $\Phi=\Phi(u)$. The EoM can then be written as
\begin{equation} 
    f\partial_u^2\Phi + f'\partial_u\Phi - \frac{u}{u_H^3}\Phi-\frac{1}{2}\Phi^3 = 0 \ .
\end{equation}
We solve this equation using standard numerical methods for ODEs; imposing regularity at the horizon leaves us with one free parameter, which can be taken to be the horizon value of the scalar field. Each horizon value yields one solution to the equation, and for each solution we extract the near-boundary coefficients $\phi^\pm$. We collect the result in the form of the source for the undeformed theory, $J=\phi^+$, as a function of the expectation value $\psi=-\phi^-$, which we plot in Fig.~\ref{fig:phi_psi}. Then, using (\ref{eq:effPotentialIntegral}), we determine the effective potential, see Fig.~\ref{fig:V_undeformed}.

We recall that we are working in units of temperature $T$. Large $\psi$-values thus correspond to low temperatures. In this limit, the effective potential must approach a conformal behavior, which, since the operator has dimension 1, implies $V(\psi)\sim \psi^3$. We have confirmed this is the case in our numerical results; more precisely, it seems to go as approximately $|\psi|^3/6$.

This potential, having only a single vacuum and being everywhere convex, is as expected too simple to provide a first-order phase transition. By including double- and triple-trace deformations we can accommodate a FOPT, as discussed next.

\subsection{Effective potential with multi-trace deformations}\label{sec:eff_potential_deformed}

The double- and triple-trace deformations, imposed through the nonlinear boundary condition (\ref{eq:nonlinearBC}), give us two new knobs $g_2$ and $g_3$ to tune. The triple-trace deformation is marginal, {\emph{i.e.}}, $g_3$ is dimensionless, while the double-trace one is relevant, $g_2$ having dimension 1. By tuning these quadratic and cubic additions to the undeformed potential as shown in Eq.~(\ref{eq:effPotDeformation}), we can find a two-vacuum structure leading to FOPTs.

\begin{figure}[t!]
\centering
\subfloat[$g_2=0$ and $g_3=0$\label{fig:V_undeformed_new_coord}]{\includegraphics[width=0.47\textwidth]{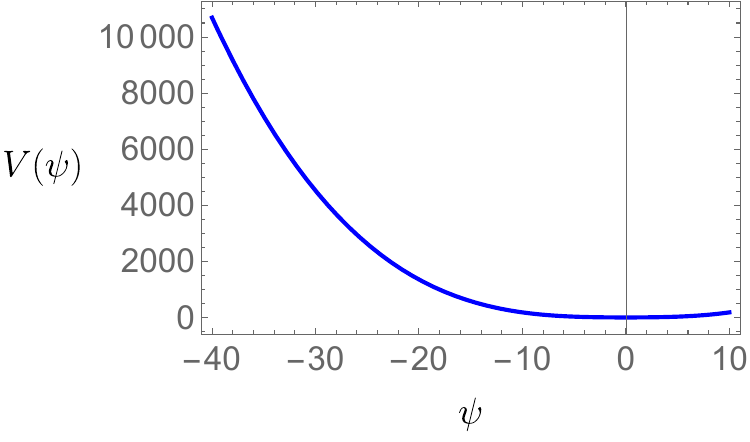}}
\qquad
\subfloat[$g_2=0$ and $g_3=0.49$ \label{fig:V_only_g3}]{\includegraphics[width=0.44\textwidth]{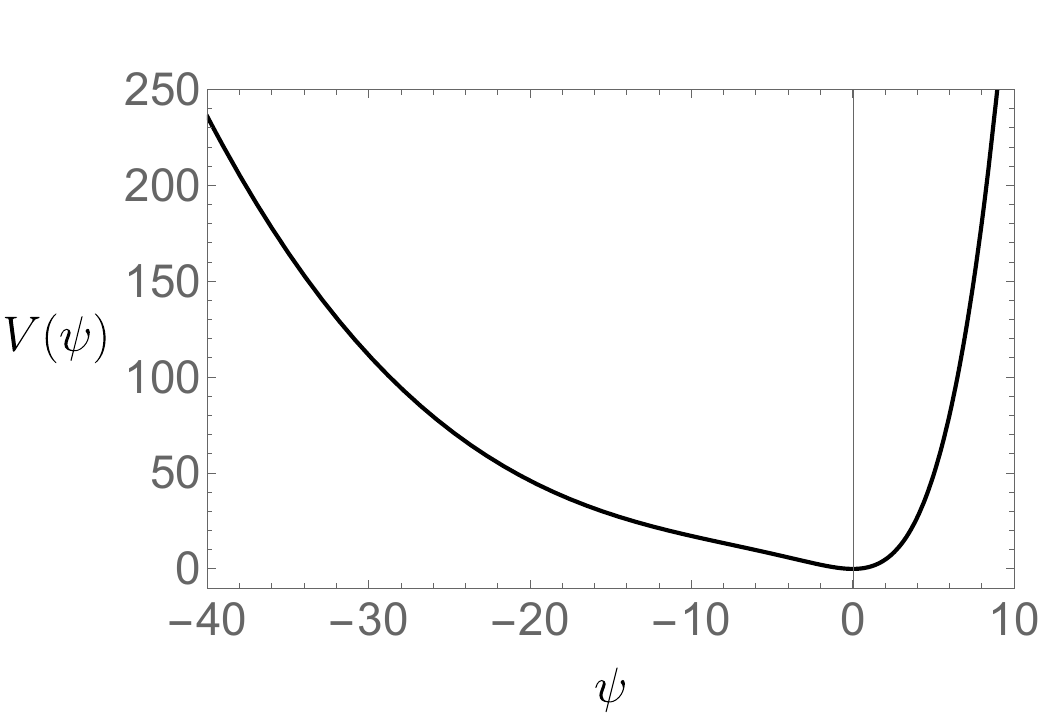}}
\caption{The effective potential $V(\psi)$ with no double-trace deformation.}
\end{figure}

First, let us consider the case when only triple-trace deformation is imposed, {\emph{i.e.}}, $g_2=0$ while $g_3\ne0$. Since the undeformed potential goes as $|\psi|^3/6$ for large values of $|\psi|$, we need $g_3<1/2$ if we want to keep the potential bounded from below. In Fig.~\ref{fig:V_only_g3}, we plot the effective potential when $g_2=0$ and $g_3=0.49$. Compared with the undeformed potential, reproduced in Fig.~\ref{fig:V_undeformed_new_coord}, we see that the value of potential in the left panel is noticeably suppressed. However, there exists only one minimum on this curve and thus no FOPT, so we need to turn on the double trace deformation as well, {\emph{i.e.}}, we need both $g_2\ne0$ and $g_3\ne0$.

\begin{figure}
\centering
\subfloat[$g_3=0.49$ and $g_2\ne 0$ \label{fig:V_PT}]{\includegraphics[width=0.45\textwidth]{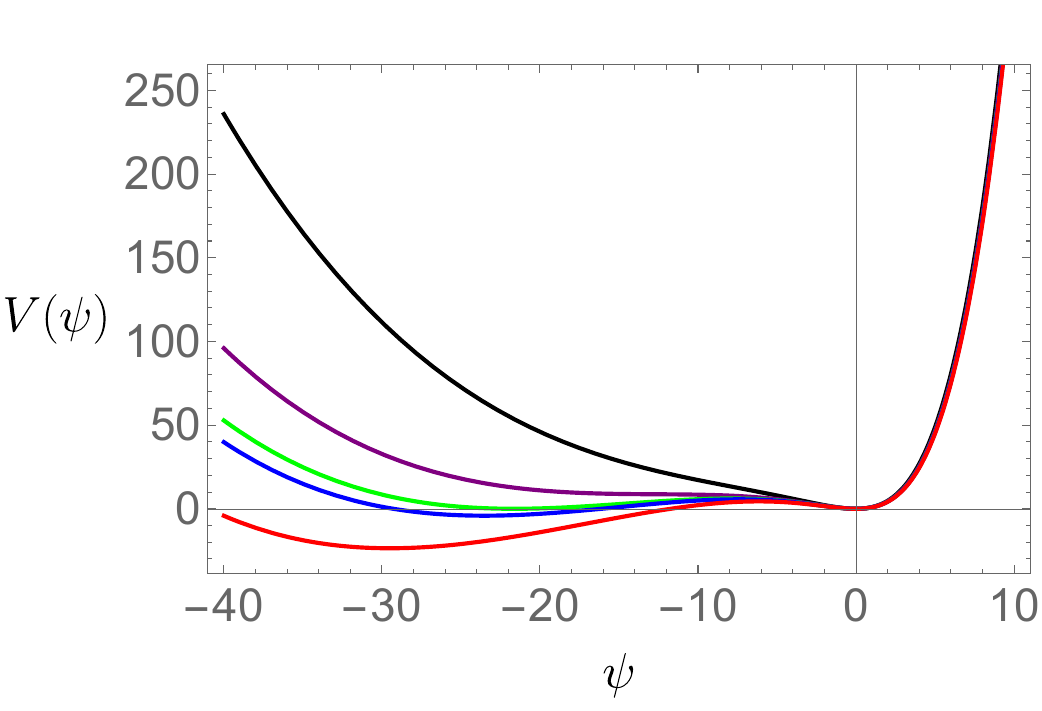}}
\qquad
\subfloat[$g_3=0.49$ and $g_2=-0.245$ \label{fig:V_0.49_-0.245}]{\includegraphics[width=0.44\textwidth]{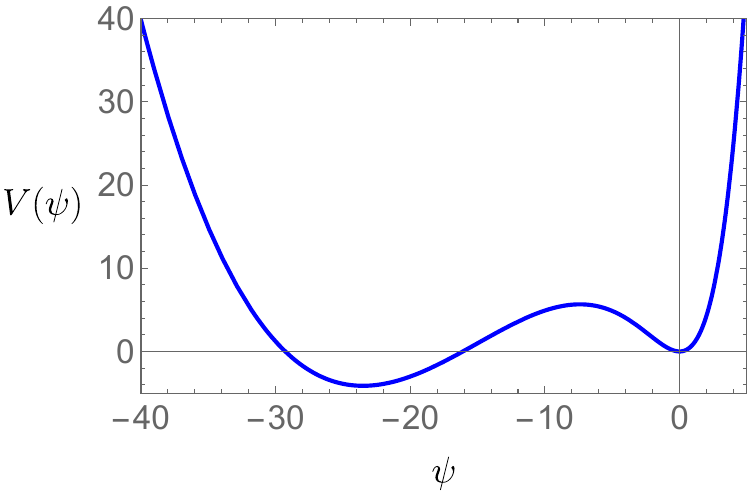}}
\caption{The effective potential $V$ plotted as a function of $\psi$ with different values of $g_2$ (see text for details).}
\end{figure}

We thus attempt to fix $g_3=0.49$ and decrease the value of $g_2$ gradually. When $g_2>-0.1794$ (in units of $T$), the potential still only has the one minimum at $\psi=0$. Then when $g_2$ drops into the interval $-0.2288<g_2<-0.1794$, there are three extreme values with two the local minimum and one local maximum. In this interval, $\psi=0$ is the global minimum, i.e., the stable state, while the other local minimum corresponds to a meta-stable state. If we  decrease the value of $g_2$ further, $\psi=0$ is no longer the global minimum when $g_2<-0.2288$.  In Fig.~\ref{fig:V_PT}, we plot the effective potential when $g_3=0.49$ and $g_2=0,-0.1794,-0.2288,-0.245,-0.3$ from top to bottom curves, respectively. In order to see its structure clearly, the curve when $g_3=0.49$ and $g_2=-0.245$ is plotted separately in Fig.~\ref{fig:V_0.49_-0.245}; there, stable, unstable, and metastable states are located at $\psi=-23.5$, $-7.4$, and $0$, respectively.

\begin{figure}[t!]
\centering
\includegraphics[width=0.45\textwidth]{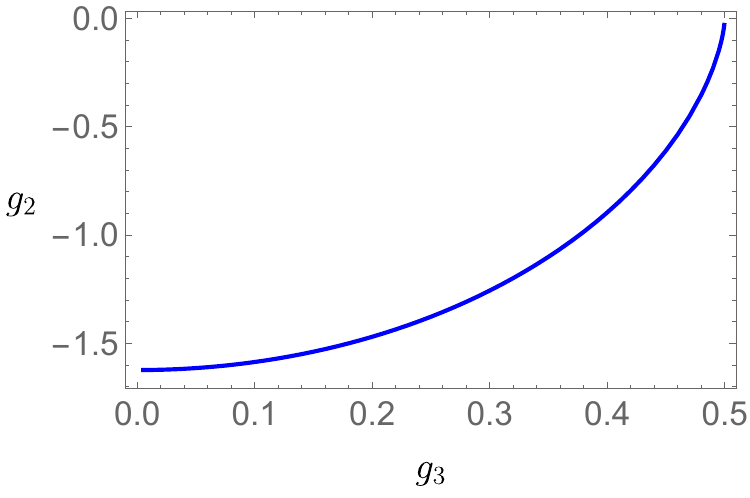}
\caption{\label{fig:phase_transition} The critical value of the double-trace coupling $g_2$ where the FOPT occurs, plotted as a function of $g_3$.}
\end{figure}

From the above results, one can conclude that when $g_3=0.49$, there is a FOPT at $g_2=-0.2288$, when the values of the potential at the two extreme minima are equal. If we also vary $g_3$, we can find the value of $g_2$ for the phase transitions as a function of $g_3$; this result is shown in Fig.~\ref{fig:phase_transition}. In this plot, $g_2$ rises 
monotonically with the increase of $g_3$, while the slope of the curve grows gradually and diverges at $g_3=0.5$. In the other end, at $g_3=0$, the curve ends in a critical point where the transition is continuous. This result is qualitatively similar to Fig.~6 of \cite{Ares:2021nap}.

\subsection{Determining the kinetic term}\label{sec:kineticTerm}

Next, we wish to determine the non-canonical kinetic term $Z(\psi)$. This can be accomplished by expanding Eq.~(\ref{eq:quantumEoM}) around a static and homogeneous solution $\bar\psi$, which is a (possibly unstable) solution to the EoM, meaning that $V'(\bar\psi)=0$. Writing $\psi(x)=\bar\psi+\xi(x)$ and neglecting terms of second order or higher in $\xi$, we find
\begin{equation}
    J(x) = \left( V''(\bar\psi) - Z(\bar\psi)\partial_\mu\partial^\mu \right) \xi(x) + \ldots \ ,
\end{equation}
which in momentum space becomes
\begin{equation}\label{eq:expandedSourceMomentum}
    \tilde J(k) = \left( V''(\bar\psi) + Z(\bar\psi)k^2 \right) \tilde\xi(k) + \ldots \ .
\end{equation}
To make contact with the dual gravity theory, we start from the expression for the source in terms of the boundary falloff of the dual field $\phi$, Eq.~(\ref{eq:nonlinearBC}):
\begin{equation}
    J(x) = \phi^+(x) + g_2\phi^-(x) - g_3\phi^-(x)^2 \ .
\end{equation}
Expanding this in the same way, we write $\phi^\pm(x)=\bar\phi^\pm+\delta\phi^\pm(x)$, where $\bar\phi$ is a homogeneous solution satisfying $\bar\phi^+ + g_2\bar\phi^- - g_3(\bar\phi^-)^2=0$. Expanding to first order in $\delta\phi^\pm(x)$ gives
\begin{equation}\label{eq:expandingHoloSource}
    J(x) = \left( \frac{\delta\phi^+(x)}{\delta\phi^-(x)} + g_2 - 2g_3\bar\phi^- \right)\delta\phi^-(x) = -\left( \frac{\delta\phi^+(x)}{\delta\phi^-(x)} + g_2 - 2g_3\bar\phi^- \right)\xi(x) \ ,
\end{equation}
where we identified $\delta\phi^-(x)=-\xi(x)$. We now go to momentum space and expand for small momenta:
\begin{equation}
   \delta\phi^\pm(k)=\delta\phi^\pm_{0} + \delta\phi^\pm_{2}k^2+\ldots \ .
\end{equation}
Inserting this into (\ref{eq:expandingHoloSource}) and comparing with (\ref{eq:expandedSourceMomentum}), we find
\begin{align}
    V''(\psi)&=-\frac{\delta\phi_0^+}{\delta\phi_0^-} - g_2 + 2g_3\phi^- \label{eq:formula_V_eff}\\
    Z(\psi)&=\frac{\delta\phi_0^{+}\delta\phi_2^{-} - \delta\phi_2^{+}\delta\phi_0^{-}}{\delta\phi_0^{-2}} \ .\label{eq:formula_Z}
\end{align}
The first of these equations gives another way to determine $V(\psi)$, which of course should be consistent with the one previously outlined. The second of these equations can be used to determine $Z(\psi)$; for each homogeneous background with scalar expectation value $\psi=-\phi^-$, we have to perturb around it in a small-momentum expansion, allowing us to determine $\delta\phi^\pm_{0,2}$.

\begin{figure}[t!]
	\centering
    \subfloat[$Z$ as a function of $\psi$. \label{fig:kinetic_term}]{\includegraphics[width=0.45\textwidth]{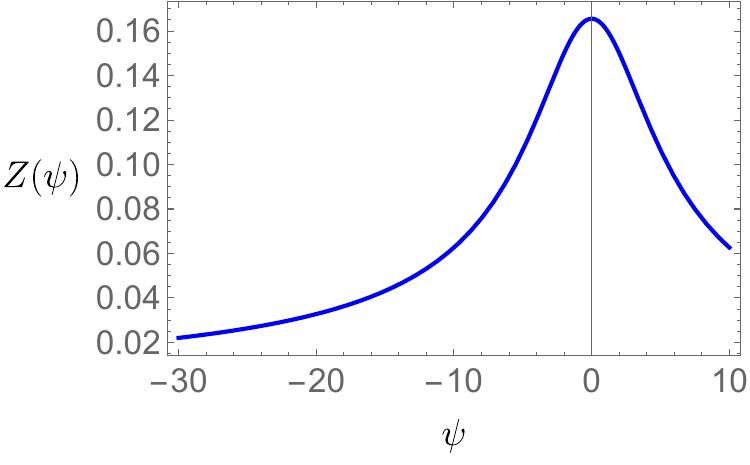}}
    \qquad
    \subfloat[Example profile of $\psi(\rho)$.\label{fig:Effective_bubble_-0.245_0.49}]{\includegraphics[width=0.42\textwidth]{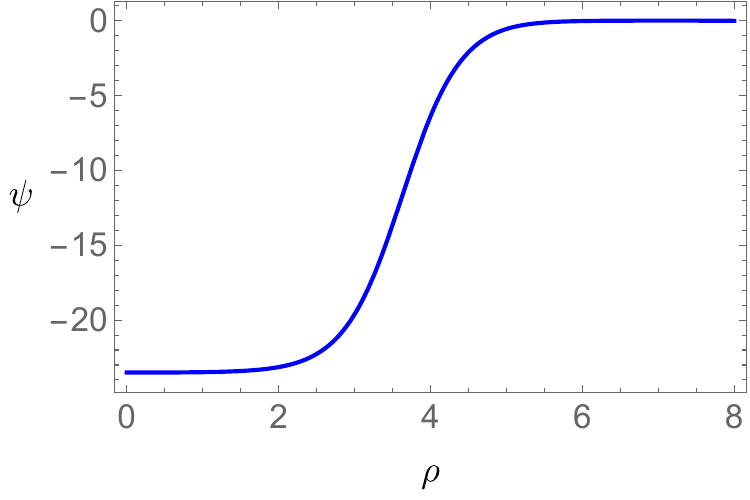}}
	\caption{The kinetic term $Z$, and the profile of the bubble solution with $g_2=-0.245$ and $g_3=0.49$.}
\end{figure}

To accomplish the above, we introduce a perturbation around the previously found homogeneous solutions,
\begin{equation}
	\Phi(u)\to \Phi(u)+\delta\Phi(u,x)\ ,
\end{equation}
where $x$ represents one of the spatial field theory directions (which one is arbitrary due to rotational invariance). Plugging this into the scalar EoM and linearizing, one obtains
\begin{equation}\label{eq:eom_delta_Phi}	f\partial_u^2\delta\Phi+f'\partial_u\delta\Phi+\partial_x^2\delta\Phi-\frac{u}{u_H^3}\delta\Phi-\frac{3}{2}\Phi^2\delta\Phi=0 \ .
\end{equation}
Next, we make a plane wave ansatz, $\delta\Phi(u,x)=e^{ikx}\delta\Phi(u,k)$, and expand for small momenta,
\begin{equation}\label{eq:Fourier_expansion}
    \delta\Phi(u,x)=e^{ikx} \left[\delta\Phi_{0}(u)+ k^{2} \delta\Phi_{2}(u)+\ldots\right] \ .
\end{equation}
Here, terms with odd powers of $k$ have to vanish because of rotational symmetry. Substituting this into Eq.~(\ref{eq:eom_delta_Phi}) and expanding for small $k$, to zeroth and second order, we get the equations
\begin{align}
	f\delta\Phi_0''+f'\delta\Phi_0-\left(\frac{u}{u_H^3}+\frac{3}{2}\Phi^2 \right)\delta\Phi_0 &=0 \label{eq:delta_Phi_0}\\
	f\delta\Phi_2''+f'\delta\Phi_2-\left(\frac{u}{u_H^3}+\frac{3}{2}\Phi^2 \right)\delta\Phi_2-\delta\Phi_0 &=0 \ . \label{eq:delta_Phi_2}
\end{align}
These equations can be solved, imposing regularity at the horizon, using a similar numerical shooting method as before.\footnote{Technically a second boundary condition at the AdS-boundary, such as fixing the value of the source, should be imposed. However, due to the linearity of the equations, this does not affect the resulting kinetic term and is largely arbitrarily.} The near-boundary behavior of these solutions has the same form as the general solution to (\ref{eq:Phi_bdry_exp}), namely
\begin{equation}\label{eq:delta_Phi_exp}
    \delta\Phi_{i}\left(u\right)=\delta\phi_i^-+\delta\phi_i^+u+\ldots \ ,
\end{equation}
where $i=0,2,\ldots$. From this and (\ref{eq:formula_Z}) we can determine $Z(\psi)$ for each $\psi$; the result is plotted in Fig.~\ref{fig:kinetic_term}.

\subsection{Finding bubble solutions}
Having obtained the effective action (\ref{eq:effectiveAction}) up to second order in derivatives, we now use it to find static critical bubble solutions. We start from the quantum EoM (\ref{eq:quantumEoM}) with $J=0$. Since we expect the bubbles to be axially symmetric, we go to polar coordinates $(\rho,\theta)$ and assume $\psi=\psi(\rho)$, resulting in the ordinary differential equation
\begin{equation}\label{eq:eom_eff_theory}
	\frac{\mathrm{d}^2\psi}{\mathrm{d}\rho^2}+\frac{1}{\rho}\frac{\mathrm{d}\psi}{\mathrm{d}\rho}+\frac{1}{2}\frac{Z'\left(\psi\right)}{Z\left(\psi\right)}\left(\frac{\mathrm{d}\psi}{\mathrm{d}\rho}\right)^2-\frac{V'\left(\psi\right)}{Z\left(\psi\right)}=0 \ .
\end{equation}
We solve this equation again using a numerical shooting method, with the boundary conditions that it should be regular in the center, $\psi'(\rho)=0$, and it should approach the false vacuum $\psi=0$ at large $\rho$. Since $\rho=0$ is a singular point of Eq.~(\ref{eq:eom_eff_theory}), we expand $\psi(\rho)$ at $\rho=0$ and solve the resulting linear equations order by order. The resulting series solution has one free parameter, which can be taken to be $\psi(0)$. Typically this value should be close to the value of $\psi$ in the true vacuum. Choosing an educated guess based on this lets us calculate $\psi(\epsilon)$ and $\psi'(\epsilon)$, where $\epsilon\ll 1$, which provides boundary conditions for our numerical integration. Integrating out to large $\rho$, we check if the solution approaches the false vacuum; if not, we adjust the free parameter in the $\rho=0$ expansion, iterating until the boundary condition is sufficiently satisfied. 

In Fig.~\ref{fig:Effective_bubble_-0.245_0.49} we show a typical bubble solution. As expected, in the small-$\rho$ region the field approaches the stable ground state, which for these parameter values is at $\psi=-23.507$, while at larger $\rho$ it approaches the metastable state.

\section{Bubble solutions directly from the gravity theory}\label{sec:bubble_gr}

In the previous section, we obtained bubble solutions using the field theory effective action. By incorporating the spatial dependence and directly solving the partial differential equation in the gravity theory, we can also obtain such bubble solutions, which will be the aim of this section. As in the previous section, we will be working in the probe limit for the scalar field, meaning that the background is fixed to pure AdS-Schwarzschild.

Due to the axial symmetry, we again find it advantageous to use polar coordinates with radial coordinate $\rho$ and with the angular variable $\theta\in\{0,2\pi\}$, meaning the metric takes the form
\begin{equation}
	\dd s^2 =\frac{1}{u^2}\left(-f \mathrm{d} t^2-2\mathrm{d} t \mathrm{d} u+\mathrm{d}\rho^2+\rho^2\mathrm{d}\theta^2\right)\ .
\end{equation}
The matter field is taken to depend only on the holographic coordinate $u$ and the radius $\rho$, $\Phi=\Phi(u,\rho)$. Thus, the scalar EoM (\ref{eq:scalarEoM}) takes on the following form:
\begin{equation}
	f \partial_u^2 \Phi+f^{\prime} \partial_u \Phi+\partial_\rho^2 \Phi+\frac{1}{\rho} \partial_\rho \Phi-\frac{u}{u_H^3} \Phi-\frac{1}{2} \Phi^3=0\ .
\end{equation}
To find critical bubble solutions, we impose the following boundary conditions:
\begin{equation}	
	\left.\partial_u \Phi-g_2 \Phi+g_3 \Phi^2\right|_{u=0}=0\ ,\ ~
	\left.\partial_\rho\Phi\right|_{\rho=0}=0\ ,\ ~
	\left.\partial_\rho\Phi\right|_{\rho=\infty}=0\ .
\end{equation}
Here, the first boundary condition imposes zero source $J=0$ in the field theory following (\ref{eq:nonlinearBC}). The second condition at $\rho=0$ imposes regularity at the center of the bubble. The third condition says that the field should return to a uniform solution (corresponding to the metastable state) infinitely far from the center of the bubble.\footnote{Usually, one would impose a Dirichlet condition, $\Phi|_{\rho=\infty}=0$, instead of the third Neumann condition, but we find that the Neumann condition is technically easier to implement numerically. For this to be valid, we must of course check that the field really does approach the metastable state at large $\rho$.} To find numerical solutions, we cut off the radial coordinate $\rho$ at some value $R_0$, chosen to be large enough to comfortably house the critical bubble, and impose the third boundary condition there instead of at infinity.

We employ pseudo-spectral methods to discretize fields in both $u$ and $\rho$ directions and solve the PDE with a Newton--Raphson method. A detailed illustration of solving PDEs with these techniques can be found in \cite{Dias:2015nua}. The additional, specialized numerical techniques used here are discussed in App.~\ref{app:numerics}. 

\begin{figure}[t!]
	\centering	
	\subfloat[The profile of $\Phi(u,\rho).$\label{fig:GR_solution_Phi_-0.245_0.49}]{\includegraphics[width=0.45\textwidth]{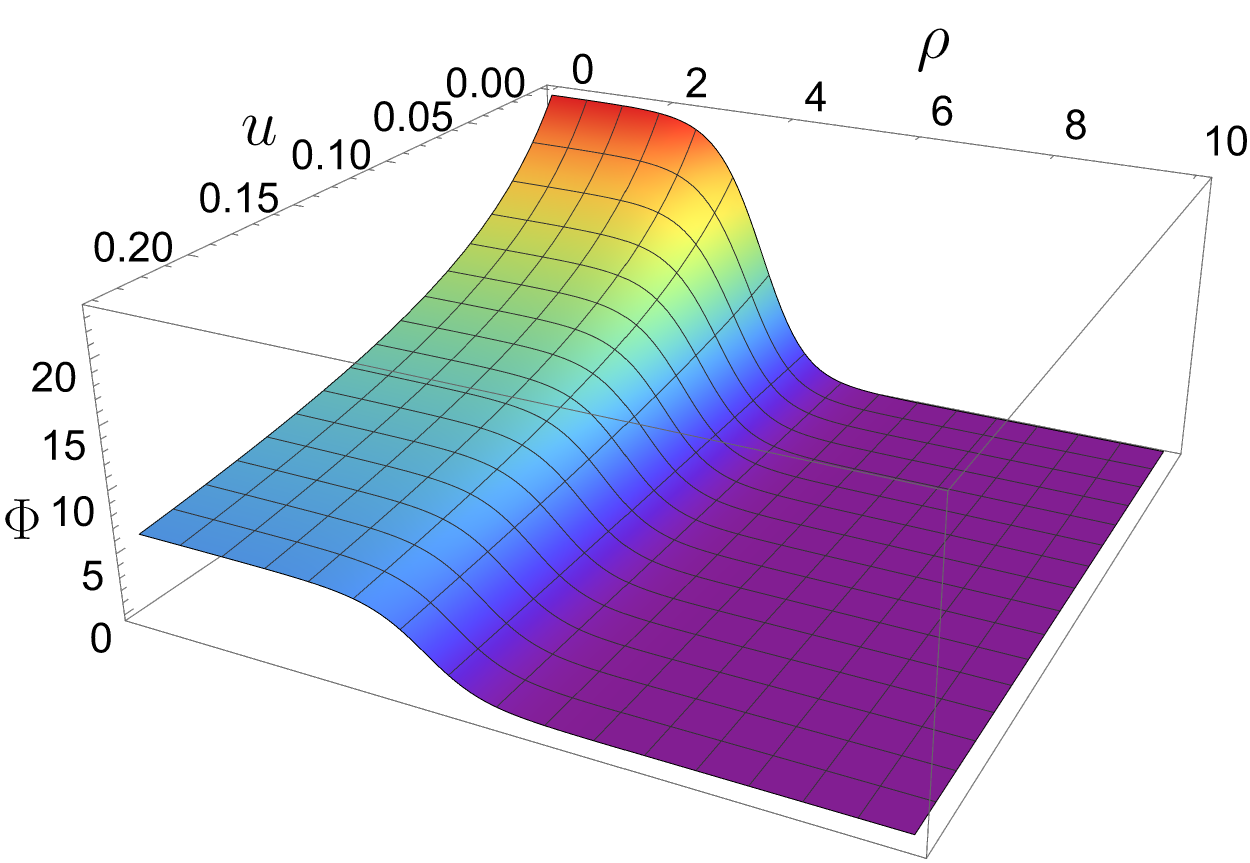}}
	\qquad
    \subfloat[\label{fig:GR_bubble_-0.245_0.49}The profile of $\psi_g(\rho)$.]{\includegraphics[width=0.45\textwidth]{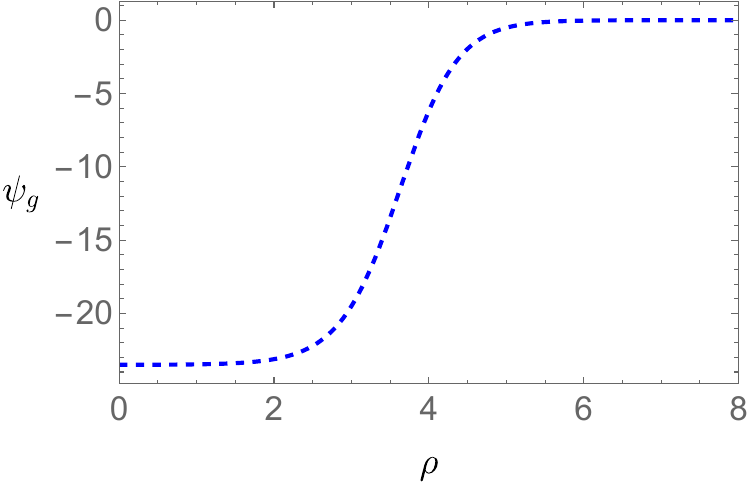}}
	\caption{The solution of the gravity theory with $g_2=-0.245$ and $g_3=0.49$.\label{fig:GR_full_-0.245_0.49}}
\end{figure}

An example solution is presented in Fig.~\ref{fig:GR_full_-0.245_0.49}. Specifically, Fig.~\ref{fig:GR_solution_Phi_-0.245_0.49} shows $\Phi(u,\rho)$, whereas Fig.~\ref{fig:GR_bubble_-0.245_0.49} shows the field theory order parameter $\psi_g(\rho)=-\phi^-(\rho)=-\Phi(u=0,\rho)$. Furthermore, the distribution of the order parameter  displayed in Fig.~\ref{fig:GR_bubble_-0.245_0.49} confirms that the obtained solution indeed represents a bubble solution. Comparison between the value of the order parameter $\psi_g(\rho)$ presented here and its value within the potential in Fig.~\ref{fig:V_0.49_-0.245}, verifies that this solution corresponds precisely to the one dictated by the given potential.

\section{Comparing the two approaches}\label{sec:comparison}

In Sec.~\ref{sec:bubble_eff} and Sec.~\ref{sec:bubble_gr}, we obtained bubble solutions from effective theory and gravity theory, respectively, and plotted examples of solutions in Fig.~\ref{fig:Effective_bubble_-0.245_0.49} and Fig.~\ref{fig:GR_bubble_-0.245_0.49}. In this section, we will compare these results.

\begin{figure}[t!]
	\centering
	\includegraphics[width=0.55\textwidth]{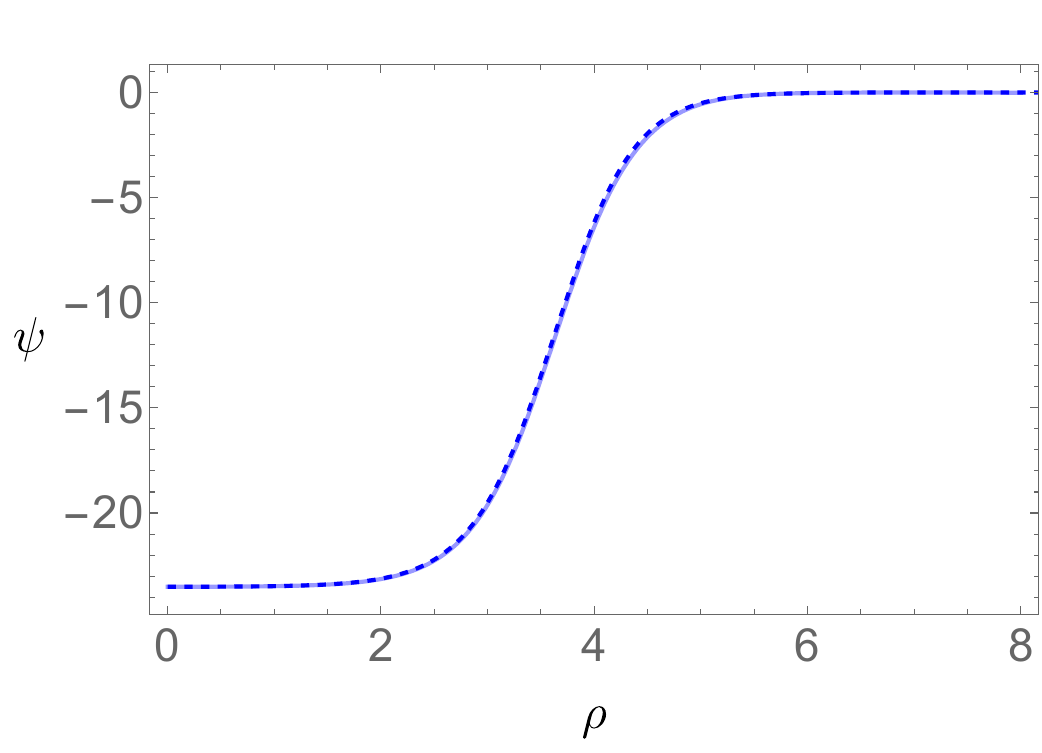}
	\caption{\label{fig:Two_bubble_solutions} Two bubble solutions plotted in one graph. In this plot, the solid curve is the bubble solution obtained from effective theory and the (overlapping) dashed curve is the solution from gravity theory.}
\end{figure}

The most straightforward way to compare the solutions is to plot them together in the same diagram; an example is shown in Fig.~\ref{fig:Two_bubble_solutions}. In this plot, the solid curve represents the bubble solution obtained from the effective theory, while the dashed curve represents the bubble solution obtained from the gravitational theory. It is clear that these two curves almost completely overlap, indicating that the solutions obtained from the two theories exhibit a very small discrepancy.

To more systematically compare the results, we pick a representative range of the parameters $g_2$ and $g_3$, as shown in Table~\ref{tb:diff_g32_ratio}. For each choice of parameters, we display the ratio of the bubble radius computed from the effective action, labeled $R_\text{EA}$, to the radius computed directly from the gravity theory, labeled $R_\text{G}$. We define the bubble radius as the distance from the origin to the point where the absolute value of $\psi$ decreases to half of its maximum. Despite the fact that the radius varies significantly as the parameters change, the ratio of the two radii stays close to 1.

\begin{table}[t!]
\centering
\begin{subtable}{0.4\textwidth}
\centering
\begin{tabular}{|c c c c|} 
    \hline
	$g_3$ & $g_2$ &  $R_\text{EA}/R_\text{G}$ & $\Delta E/\Delta V$ \\ [0.5ex]
    \hline\hline
    0.49 & $-0.234$  & 1.004 & 0.209\\
    \hline
    0.49 & $-0.238$  & 1.004 & 0.385\\
    \hline
    0.49 & $-0.242$ & 1.004 & 0.576\\
    \hline
    0.49 & $-0.246$  & 1.003 & 0.782\\
    \hline
    0.49 & $-0.250$ & 1.003 & 1.003\\
    \hline
    0.49 & $-0.400$  & 1.004 & 28.652\\
    \hline
	\end{tabular}
\caption{Solutions with fixed $g_3$.}
\end{subtable}
\hspace{0.075\textwidth}
\begin{subtable}{0.4\textwidth}
\centering
\begin{tabular}{|c c c c|} 
    \hline
	$g_3$ & $g_2$ & $R_\text{EA}/R_\text{G}$ & $\Delta E/\Delta V$ \\ [0.45ex]
    \hline\hline
    0.49 & $-0.245$ & 1.003 & 0.729\\
    \hline
    0.48 & $-0.380$ & 1.003 & 1.096\\
    \hline
    0.47 & $-0.480$ & 1.003 & 1.071\\
    \hline
    0.46 & $-0.560$ & 1.004 & 0.860\\
    \hline
    0.45 & $-0.640$ & 1.004 & 1.088\\
    \hline
    0.44 & $-0.700$ & 1.004 & 0.804\\
    \hline
	\end{tabular}
\caption{Solutions with variable $g_2$ and $g_3$.}
\end{subtable}%
\caption{\label{tb:diff_g32_ratio} Comparison of solutions using the two different approaches. The ratio of bubble radii, with $R_\text{EA}$ and $R_\text{G}$ corresponding to the effective theory and gravitational solutions, respectively, match remarkably well for a wide range of $\Delta E/\Delta V$.}
\end{table}

The form of the critical bubbles depend to a high degree on how large the supercooling is, \textit{i.e.}, how close are the two competing vacua to being degenerate? Near degeneracy, we are in the so-called thin-wall limit \cite{Coleman:1977py}. There, the bubbles can be cleanly divided into an outside consisting of the metastable phase, an inside consisting of the stable phase, and an interpolating bubble wall whose thickness is small compared to the total bubble size. On the other hand, for large supercooling we typically approach a spinodal point, where the barrier between the two vacua disappears completely. In this \emph{thick-wall} limit, the critical bubble cannot be cleanly separated into an inside and an outside, and the value of the order parameter in the center of the bubble does not get close to the true vacuum. In order to quantify the amount of supercooling, we compute, for each set of multi-trace deformations in Table \ref{tb:diff_g32_ratio}, the ratio $\Delta E/\Delta V$, with $\Delta E$ being the difference in the effective potential between the false and true vacua, and $\Delta V$ the height of the potential barrier measured from the false vacuum. For low supercooling, near the critical point, $\Delta E/\Delta V$ approaches zero, while for large supercooling, close to spinodal point, it diverges. From the values in the tables, we can see that our test cases include both small, intermediate, and large values of $\Delta E/\Delta V$.

\begin{figure}
	\centering
	\includegraphics[width=0.7\textwidth]{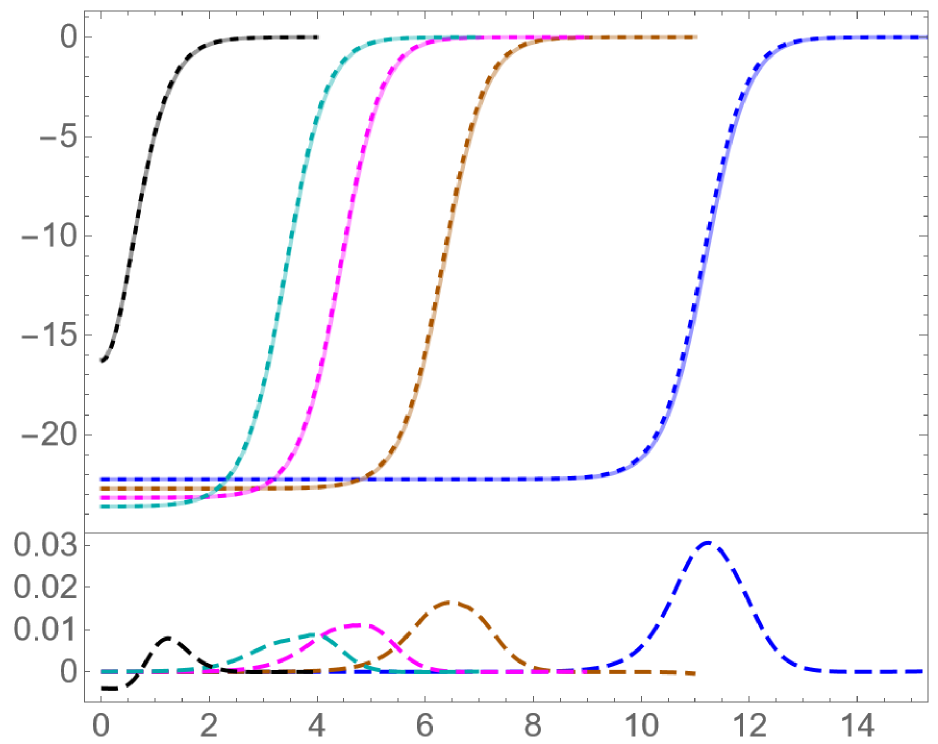}
	\caption{\label{fig:profiles_comparison} Profiles of critical bubbles from both solution methods, as well as their relative deviations, as functions of the field theory radial coordinate $\rho$. The upper part of the figure illustrates the configuration of the bubble solution, where the lighter solid lines represent the solutions from the effective theory, while the dashed lines correspond to the solutions from the associated gravitational theory. The lower part of the figure depicts the relative deviation $\delta(\rho)$, defined in (\ref{eq:difference}). In all five cases shown, $g_3=0.49$, while the blue, orange, magenta, cyan, and black curves correspond to $g_2 =-0.234, -0.238,-0.242,-0.246$, and $-0.4$, respectively.}
\end{figure}

To achieve a more detailed comparison between the two bubble solutions, we can compute the difference between the solutions,
\begin{equation}\label{eq:difference}
    \delta(\rho)\equiv\frac{\psi_\text{EA}(\rho)-\psi_\text{G}(\rho)}{\psi_\text{EA}(\rho=0)} \ ,
\end{equation}
where $\psi_\text{EA}(\rho)$ and $\psi_\text{G}(\rho)$ are solutions using the effective action and gravity approach, respectively, and we normalize by the value of the effective action solution at the bubble center. In Fig.~\ref{fig:profiles_comparison}, we fix $g_3=0.49$  and present the profiles of different bubbles, ranging from the thick-wall limit to the thin-wall limit, along with the corresponding $\delta(\rho)$. In the upper part of the figure, solid and dashed lines represent the bubble solutions from the effective theory and gravitational theory, respectively. The two curves align closely, confirming the consistency between the results of both theories. The lower part of the figure shows the spatial distribution of $\delta(\rho)$, highlighting the differences between the two solutions. We observe that the difference is practically zero at the bubble center, increases to a maximum of a few percent at the bubble wall, and then decreases again. Moreover, as we get closer to the thin-wall limit, the maximum value of $\delta(\rho)$ gradually increases. 

A physically important measure of the difference of the solutions is the difference in the resulting decay rates, which goes as $e^{-S_\text{B}}$ with $S_\text{B}$ the action evaluated on the bubble solution. Evaluating the two-derivative effective action on the two different solutions and computing $e^{-S_\text{B}}$, we find deviations of 1--3 \%.

Lastly, we note that domain wall solutions, towards which the bubble solutions tend in the thin-wall limit, can in many cases be well-fitted by a hyperbolic tangent \cite{Janik:2021jbq,Ares:2021ntv}. With this in mind, we have fitted some of our bubble profiles to a function of the form $A\left( \tanh(k\, \rho-\delta) - 1 \right)$, with $A$, $k$, and $\delta$ as fitting parameters. For bubbles with $\Delta E/\Delta V$ small (thin-wall limit) or near 1, we find very good agreement, with errors well below the percent level. With $\Delta E/\Delta V$ large, on the other hand, we find errors of around five percent. It should be emphasized that only in the extreme thin-wall limit does the fitting function respect the boundary condition $\psi'(\rho=0)=0$.

\section{Discussion}\label{sec:discussion}

We studied a simple Einstein-scalar system, acting as a bottom-up holographic dual of a strongly coupled QFT. Working in the probe approximation and introducing a multi-trace deformation at the boundary, we found that the theory exhibits a rich phase diagram including FOPTs. We then constructed critical bubble solutions, which describe how the system transitions across such a FOPT, using two distinct approaches: 
\begin{enumerate}[(I)]
 \item By solving the scalar field equation with full spatial dependence in the black brane background in Sec.~\ref{sec:bubble_gr}, and
 \item By constructing a two-derivative effective action for the order parameter and solving the resulting boundary field equation in Sec.~\ref{sec:bubble_eff}.
 \end{enumerate}
The two methods differ substantially in complexity. The direct gravitational construction (I) requires solving a nonlinear PDE with nontrivial boundary conditions. The effective action approach (II), by contrast, involves only ODEs, both in its construction (from homogeneous and linearized bulk solutions) and in the computation of the bubble profiles. Despite this difference, we found excellent agreement between the resulting bubble solutions across the parameter range studied. This confirms that, at least for the static, spherically symmetric configurations in this paper, the derivative expansion provides a quantitatively reliable description.

Although these are promising results, it would be prudent to extend the tests to a larger variety of systems and scenarios. Perhaps most importantly, the possible influence of backreaction of the scalar field should be studied. We cannot exclude the possibility that backreaction will increase the discrepancy between the two approaches.\footnote{However, we note that in \cite{Ares:2021ntv}, which did include backreaction, some higher-derivative corrections to the effective action were computed and found to be small when evaluated on bubble solutions.} Other dimensions, as well as more complicated gravitational Lagrangians, would also provide good checks. The inclusion of higher-derivative terms in the effective action would be an important complement. There is also the question of exactly \emph{which} effective action to employ; this is particularly important if one wants to compute nucleation rates with increased precision, necessitating the inclusion of fluctuations around the bubble saddle point. Alternatives to the derivative expansion of the full 1PI action which we have studied here are various course-grained or high-temperature effective actions (see \cite{Strumia:1998nf,Gould:2021ccf,Kierkla:2025qyz,Navarrete:2025yxy} for examples and more details). It would be interesting to better understand their holographic duals. Regardless of these concerns, however, the results herein show that the two-derivative quantum effective action can successfully capture the details of the corresponding classical gravity solutions, which is already an important accomplishment.

Encouraged by this we propose that the effective action method (II) be extended to a variety of more general situations beyond bubble nucleation. One natural direction is the study of other inhomogeneous configurations such as vortices, domain walls, or spatially modulated phases. Setups involving a spatially modulated source, such as holographic lattices~\cite{Horowitz:2012ky}, could be conveniently studied by solving the EoM resulting from the effective action with a suitable source term. Spontaneous translational symmetry breaking can perhaps also be obtained using this framework. This could be caused by the function $Z(\psi)$ in the kinetic term changing sign, although higher order derivative terms might be needed to see such an instability or to construct the resulting end state.

Holographic duality has been fruitfully applied to the study of non-zero density QCD matter, relevant in the context of neutron stars~\cite{Hoyos:2016zke} (see also  reviews~\cite{Hoyos:2021uff,Jarvinen:2021jbd}). At non-zero density, there may exist a FOPT between nuclear and quark matter (possibly in some paired Higgs phase) \cite{Ecker:2025vnb}. The effective action approach could be employed in holographic models with such transitions to compute quantities such as nucleation rates and surface tensions between the two phases.

By extending the effective action to include time derivatives, one could use it to study real-time phenomena, which is another realm where holography shows its strength. This includes spinodal decomposition \cite{Attems:2017ezz,Janik:2017ykj,Attems:2019yqn,Caddeo:2024lfk,Chen:2024pyy} and bubble wall propagation \cite{Bigazzi:2021ucw,Bea:2021zsu,Henriksson:2021zei,Janik:2022wsx,Henriksson:2024hsm}. In the latter case, one may be able to track the bubble wall velocity and its eventual saturation under friction-like effects without resorting to full numerical solution of the Einstein equations.

Finally, including the stress tensor in the effective action may open up further directions, allowing the study of transitions in time-dependent or curved boundary geometries~\cite{Compere:2008us,Ishibashi:2023luz}. This could connect to problems involving dynamical backgrounds such as those relevant to cosmology \cite{Ecker:2021cvz,Ecker:2023uea}.

Altogether, our results provide a controlled test of the effective action approach to inhomogeneous physics in holographic theories. They suggest that the method, while approximate, captures features of non-perturbative processes such as bubble nucleation with good accuracy, and may be fruitfully applied to a broader range of strongly couple phenomena.

\vspace{1.0cm}
\begin{acknowledgments}
We thank Carlos Hoyos, Javier G. Subils, and Hongbao Zhang for useful discussions. O.~H. has been supported in part by the Waldemar von Frenckell foundation. N.~J. has been supported in part by the Research Council of Finland grant no.~3545331. N.~J. wishes to thank the organizers of the ``Holographic applications: from Quantum Realms to the Big Bang'' conference for kind hospitality while this work was being finished. X.~L. acknowledges the support form China Scholarship Council (CSC, No. 202008610238).
\end{acknowledgments}

\appendix

\section{Holographic renormalization}\label{app:holorenorm}

Here we give some details on the holographic renormalization of our gravity theory. We start from the action in (\ref{eq:bulkAction}) and focus on the scalar part (neglecting the factor of $1/N$ for brevity):
\begin{equation}\label{eq:bulkActionScalar}
S_\phi=-\frac{1}{2\kappa_{4}^{2}}\int\mathrm{d}^{4}x\sqrt{-g}\left[\partial_{\mu}\phi\partial^{\mu}\phi+m^{2}\phi^{2}+\frac{1}{4}\phi^{4}\right].
\end{equation}
Asymptotically the metric approaches AdS$_4$,
\begin{equation}
    ds^2 \to \frac{1}{u^2}\left(-\mathrm{d} t^2+\mathrm{d} u^2+\mathrm{d} \vec{x}^2\right) 
\end{equation}
and the scalar field goes as
\begin{equation}
    \phi(x) \to \phi^-(t,\vec x)\, u + \phi^+(t,\vec x)\, u^2 \ .
\end{equation}
We cutoff the asymptotically AdS-space at some radius $u=u_0$ near the boundary. For the variation of the action to be finite on solutions, we need to add a set of counterterms living on the cutoff surface,
\begin{equation}
    S_{CT} = \frac{1}{\kappa_4^2}\int_{u=u_0}\mathrm{d}^{3}x\sqrt{\gamma}\left[ c_1\phi \, n^\mu \partial_\mu\phi + c_2\phi^2 \right] \ .
\end{equation}
Here, $n^\mu$ is the unit normal to the cutoff surface, and $\gamma$ is the determinant of the induced metric on the cutoff surface, which near the boundary equals $u^{-3}$. Choosing the coefficients to obey $2c_2=2c_1-1$ removes all divergences in the variation of the action, resulting in
\begin{equation}
    \delta_\phi (S_\phi + S_{CT}) = \frac{1}{\kappa_4^2}\int_{u=u_0}\mathrm{d}^{3}x\sqrt{\gamma}\left[ -\phi^+(c_1-1)\delta\phi^- - c_1\phi^-\delta\phi^+ \right] \ .
\end{equation}
Furthermore setting $c_1=1$ puts us in alternate quantization, with the action vanishing on variations if the source $\phi^+$ is kept constant, and with the expectation value of the field theory operator $\mathcal{O}$ dual to $\phi$ being
\begin{equation}
    \langle \mathcal{O} \rangle \equiv \kappa_4^2\frac{\delta S}{\delta\phi^+} = -\phi^- \ .
\end{equation}
We can also add another, non-covariant term to the cutoff surface action:
\begin{equation}
    S_W = \frac{1}{\kappa_4^2}\int_{u=u_0}\mathrm{d}^{3}x\sqrt{\gamma}\left[ -\phi^- W'(-\phi^-)-W(-\phi^-) \right] \ .
\end{equation}
The point of this new term is that it lets us modify the boundary condition for the bulk field; in the dual field theory, it corresponds to implementing a deformation by an operator $W(\mathcal{O})$ \cite{Papadimitriou:2007sj}. Including this term, the variation of the total action $S_\text{tot}=S_\phi + S_{CT}+S_W$ becomes
\begin{equation}\label{eq:totalActionVariation}
    \delta_\phi S_\text{tot} = \frac{1}{\kappa_4^2}\int_{u=u_0}\mathrm{d}^{3}x\sqrt{\gamma}\left[ -\phi^-\delta(\phi^+ + W'(-\phi^-) \right] \ .
\end{equation}
We will choose the following $W$, corresponding to a deformation by a quadratic and a cubic term:
\begin{equation}
    W(-\phi^-) = \frac{g_2}{2}(-\phi^-)^2 + \frac{g_3}{3}(-\phi^-)^3 \ .
\end{equation}
Then, Eq. (\ref{eq:totalActionVariation}) tells us that in order for the variation of the action to be zero, we must hold fixed the quantity
\begin{equation}\label{eq:boundaryCondition}
    J = \phi^+ + g_2\phi^- - g_3(\phi^-)^2 \ .
\end{equation}
As per the usual AdS/CFT dictionary, we can interpret this $J$ as a source (or a coupling) for the single-trace operator $\mathcal{O}$, and we still have $\langle \mathcal{O} \rangle=\kappa_4^2\frac{\delta S_\text{tot}}{\delta J}=-\phi^-$. Throughout the paper, however, we have considered the case with vanishing single-trace coupling, $J=0$, and studied the theory as a function of the double- and triple-trace couplings $g_2$ and $g_3$.

\section{Details of the numerics}\label{app:numerics}
For a detailed explanation of how to solve (partial) differential equations using the Newton--Raphson method, see~\cite{Dias:2015nua,Krikun:2018ufr}. In this appendix, we primarily focus on the specific challenges we encountered in solving the PDE in Sec.~\ref{sec:bubble_gr}.

Although the PDE under consideration is relatively simple, the difficulty arises from the mixed nonlinear boundary condition
\begin{equation}
    \left.\partial_u \Phi-g_2 \Phi+g_3 \Phi^2\right|_{u=0}=0 \ .
\end{equation}
Compared with standard Dirichlet or Neumann boundary conditions, this one is more complicated, which makes the Newton--Raphson method particularly susceptible to yielding trivial solutions. 

Typically, the key to successfully solving equations using the Newton--Raphson method lies in selecting an appropriate initial guess. We found that when  $u_H=1$ and the number of sampling points is relatively small, non-trivial solutions are easier to obtain. Under these circumstances, we first obtain a solution, which is then used as the initial guess for further iterations. Subsequently, we make small adjustments to parameters $u_H$, $g_2$, $g_3$, and the number of sampling points to obtain new solutions. These new solutions are then used as initial guesses for the next iteration.

At this point, further explanation is needed on how to modify the number of sampling points in the numerical solution. We use the Chebyshev spectral method to expand the field, and the solution we obtain represents the values of the field at different locations under the current sampling points. To adjust the number of sampling points, we simply substitute these function values back into the expansion and then invert the Chebyshev matrix:
\begin{equation}\label{eq:cheb}
    f(x_m)=\sum_n C_n T_{n-1}(x_m)\equiv \sum_n C_n T_{nm} \ ,
\end{equation}
where $T_{n}(x)$ is the Chebyshev polynomials of the first kind. The coefficients $C_n$ can be obtained by inverting the series (\ref{eq:cheb}). Once we have determined the coefficients from the expansion with the current number of sampling points (say $N_0$), we can use these coefficients as the coefficients for the first $N_0$ terms of the higher-order case, while setting the higher-order coefficients to zero. Because increasing the number of sampling points, the situation is, in fact, equivalent to considering a higher-order expansion. This approach yields an approximate numerical solution, which can then be used as the initial guess for solving the case with more sampling points.


\bibliographystyle{JHEP}
\bibliography{refs}

\end{document}